# Andreev reflection observation in Nb/Fe$_{0.5}$Si$_{0.5}$/Nb and Nb/Fe$_{0.5}$Si$_{0.5}$/Si/Nb Josephson junctions

O. Vávra, Š. Gaži, I. Vávra, G. Radnóczi*, V.V. Moshchalkov**, J. Dérer and E. Kováčová

Institute of Electrical Engineering SAS, Dúbravská cesta 9, SK-841 04 Bratislava, Slovak Republic

* Research Institute for Technical Physics and Materials Science, P.O. Box 49, H-1525 Budapest, Hungary

** Katholieke Universiteit Leuven, Nanoscale Superconductivity and Magnetism Group, Laboratory for Solid State Physics and Magnetism, Celestijnenlaan 200 D, B-3001 Leuven, Belgium

## Abstract

Electrical properties of Josephson junctions Nb/Fe$_{0.5}$Si$_{0.5}$/Nb with superconductor/ferromagnet (S/F) interfaces are presented. Due to Andreev reflection the nearly exact quadruple enhancement of the tunnel junction differential conductance compared with that of the normal state was achieved. The transparency of the S/F interfaces in our junctions was estimated to be close to unity. This almost ideal value is obtained due to the use of a very smooth amorphous magnetic Fe$_{0.5}$Si$_{0.5}$ alloy for the barrier preparation. The real structure of the Nb/Fe$_{0.5}$Si$_{0.5}$/Nb tunnel junction is described as a S/F/I/F/S junction. Also Nb/Fe$_{0.5}$Si$_{0.5}$/Si/Nb Josephson junctions were investigated and the results found on these junctions confirm the effects observed in Nb/Fe$_{0.5}$Si$_{0.5}$/Nb.

# 1. Introduction

In recent years considerable attention has been devoted to the research of hybrid superconductor/ferromagnet (S/F) and superconductor/normal metal (S/N) structures. The electron can enter into the superconducting condensate and contribute to the supercurrent only if it is coupled to another electron thus forms Cooper pair. The other electron required for the formation of the pair can be taken from the normal metal, leaving behind a hole at the interface. This hole will be reflected from N/S interface with momentum opposite to that of the incident electron and it propagates away from the interface. This conversion of a normal current to a supercurrent at the S/N interface is known as the Andreev reflection (Andreev 1964). The Andreev-reflected holes act as a parallel conduction channel to the initial electron current, thus doubling the normal-state conductance $G_n$ of the S/N interface for applied voltages eV < Δ, where Δ is the superconducting gap.

The induced superconducting order parameter decays in the N region as the distance from the S/N interface increases. Therefore for a successful observation of the Andreev reflection at the S/N interfaces, the N layer must be sufficiently thick compared with the coherence length $\xi_N$ (Soulen et al 1998). To achieve a high efficiency of the Andreev reflection, a maximum transparency T of the S/N interface is needed. The transparency is defined as $T = 1/(1+Z^2)$, where Z is a dimensionless parameter proportional to the potential barrier at the interface (Blonder et al 1982).

In the case of the S/F interface electrons additionally can lose their correlation under the exchange field in the F region. Thus the coherence length $\xi_F$ is usually smaller than $\xi_N$. An exchange field is also responsible for spatial oscillations of the induced superconducting order parameter in the F-layer. The induced superconducting order parameter with negative sign is usually referred to as a π-state (Z. Radović et al 1991). For some suitable values of exchange energy and the F-layer thickness $d_F$, one can obtain S/F/S junction critical current vs. temperature, $I_C(T)$, oscillation with vanishing $I_C$ for certain values of temperature and $d_F$ (Veretennikov et al 2000). This characteristic temperature is related to the transition of the S/F/S junction from 0 to $\pi$ -state. The 0 to $\pi$ -transition at low temperatures was predicted also for the S/F/I/F/S junctions even if their F-layers thickness is smaller than $\xi_F$ (considering the following assumptions: a small transparency of the insulating (I) barrier, and a large transparency of the S/F interfaces (Golubov et al 2002)).

## 2. Sample preparation

Our aim was to prepare the junctions with a high transparency T of the S/F interfaces and a high resistance barrier. In present work we investigated the junction properties by the observation of the Andreev reflection. To obtain a high value of transparency T, it is a prerequisite to achieve a low roughness of the S/F interface and a uniform concentration of the ferromagnetic material along the S/F interface (Zareyan et al 2001). We have used magnetic amorphous $Fe_{0.5}Si_{0.5}$ alloy as the material for the Josephson tunnelling barrier. Comparing amorphous material with the polycrystalline magnetic barrier material, one should note that the amorphous state of the magnetic layer guarantees the homogeneity of the barrier composition on microscopic level. The absence of microdefects (grain boundaries, dislocations) in the barrier implies also its magnetic homogeneity.

A particular attention has been given to the structural characterization of the $Nb/Fe_{0.5}Si_{0.5}/Nb$ junctions. The crystalline structure of Nb electrodes and $Fe_{0.5}Si_{0.5}$ barrier layer were determined by the high resolution electron microscopy – HREM (Fig.1). The results were compared with structural data obtained earlier on the Nb/Si/Nb junctions prepared at very similar conditions (Vávra et al 1997).

The Nb electrodes are polycrystalline with a strong (110) preferential orientation in the normal to the film plane direction. The FeSi barrier has an amorphous structure. The FeSi barrier thickness is uniform along the whole contact area. No "crystalline bridges" in the barrier were observed. The main structural difference in comparison to Nb/Si/Nb junction is in minimizing the interlayer transition zones (ITZ) at the FeSi-on-Nb and Nb-on-FeSi interfaces. From the HREM micrograph it is clearly seen that the ITZs, if they exist, are restricted to a few atomic planes. Concerning the difference between the two ITZs it is impossible to come to any conclusion just having the HREM picture. Usually the deposition of a heavier on a lighter element involves a thicker ITZ (Yulin et al 2002). It means that the Nb-on-FeSi interface (top interface) should be thicker comparing to the FeSi-on-Nb (bottom interface). The non-equivalency of these two interfaces is registered in present work by the electrical measurements.

The Nb/Fe$_{0.5}$Si$_{0.5}$/Nb junctions were fabricated by sputtering technique at deposition conditions very similar to those used for the fabrication of the Nb/Si/Nb junctions (Vávra et al 1997). Three deposition runs were used for the junction preparation. In the first run the bottom superconducting electrode (150 nm), Fe$_{0.5}$Si$_{0.5}$ barrier (typically ~ 3 ÷ 5 nm), and a fraction of the top Nb electrode (40 nm) were deposited onto an oxidized Si wafer. These stacked layers prepared in the first vacuum run were patterned by Ar ion beam etching to obtain a tunnel junction of a size of 20 x 20 μm$^2$. The insulating layer between the bottom and the upper electrodes outside the junction area was prepared in the second deposition run by rf sputtering of SiO$_2$. The last deposition run was the sputtering of the Nb leads (approx. 150 nm) on the top Nb electrode. The lift-off technique was used for the patterning of the upper leads. The critical temperature of both electrodes, measured by a conventional resistive method, is 8.5 K and 6.28 K, respectively.

## 3. Experimental results

The fact that our junctions are tunnel-like with a pinhole-free barrier is confirmed by a temperature dependence of the zero-bias resistance R$_{TJ}$(T) (Åkerman et al 2002). Figure 2 clearly shows a non metallic R$_{TJ}$(T) dependence of the Nb/Fe$_{0.5}$Si$_{0.5}$/Nb junctions.

In the superconducting junctions with normal metal barrier the multiple Andreev reflection appears. It is usually observed as a sub-harmonic gap structure in the current-voltage curves at voltages $2\Delta/ne$, where $n$ is integer. This feature is typical for the tunnel junctions with pinholes (Kleinsasser et al 1994). As we show below our junctions do not have such feature.

Irradiation of the Nb/Fe$_{0.5}$Si$_{0.5}$/Nb tunnel junction with the microwave frequency ν=10 GHz revealed well-defined Shapiro steps at $V_n=nh\nu/(2e)$. For ν=10 GHz the steps are expected at *n* x 20.6 µV (*n* is integer), which is in good agreement with the steps clearly seen at voltages near 20, 40, 60 and 80 µV (Fig. 3b, inset). This gives direct experimental evidence for the presence of the ac Josephson effect in our tunnel junctions.

### 3.1. S/F/I/F/S junctions

Current-voltage characteristics of the Nb/Fe$_{0.5}$Si$_{0.5}$/Nb junction were measured (Fig. 3), and subsequently numeric calculations were carried out to obtain the differential conductance dI/dV vs. bias voltage curve (Fig. 4). In the subgap regime (V < 0.9 mV) an enhancement of differential conductance was observed. The broad peak is the consequence of Andreev reflection at the S/F interfaces. The efficiency of Andreev reflection *A* is characterized by the ratio of the broad peak maximum to the normal conductance $G_n$ (both parameters are marked by arrows in Fig. 4). A narrow peak at the voltages close to zero and the dip at the gap edge appear if both electrodes are superconducting.

In simple terms, due to Andreev reflection at the interfaces with the transparency $T \to 1$, one S/F interface in subgap regime exhibits a *double* enhancement of the differential conductance compared with that in the normal state (*A* = 2). Similar situation happens for the real S/N interface described by Arnold as the S/I/N junction with low effective barrier thickness (Arnold 1985). The two independent ideal S/F interfaces connected in series also exhibit *A* = 2 for voltages V < 2Δ/e.

If, however, one S/F/S tunnel junction contains two ideal interfaces S/F and F/S with $T \to 1$, the above mentioned effect of Andreev reflection is doubled for voltages *V* < Δ/e. Consequently, in subgap regime a *quadruple* enhancement of $G_n$ should be observed (*A* = 4). If the applied voltage is higher than the gap voltage of one of the S-electrodes Andreev reflection can not be observed.

Figure 5 shows the temperature evolution of the differential conductance characteristics of the Nb/Fe$_{0.5}$Si$_{0.5}$/Nb junction. As the temperature increases to the critical temperature of the upper electrode, the differential conductance dI/dV versus bias voltage changes, and the dip at the gap edge disappears. At a temperature of 6.12 K, which is close to the critical temperature of the upper electrode, superconducting tunneling current vanishes and only the bottom S/F interface exhibits Andreev reflection with $A_B \approx 2$. Since at this temperature Andreev reflection at the top interface does not occur its efficiency is $A_T \approx 1$. At temperatures lower than 6 K Andreev reflection from both S/F interfaces gives values *2.28 < A < 3.76*. At a temperature of 4.2 K the efficiency of Andreev reflection is *A ≈ 3.76*. The bottom S/F interface has again *$A_B$ ≈ 2* (normally the value not exceeding 2 is possible), and for the top S/F interface *$A_T$ ≈ 1.88 (A = $A_B$\*$A_T$)*. Thus, based on the measurements, one can deduce the transparency *T* of the bottom S/F interface close to 1, which is almost ideal. The top S/F interface has a transparency of *T < 1*.

Taking into account the experimentally observed electrical non-equivalency of the tunnel barrier interfaces and the fact that Si atoms predominantly diffuse into Nb (Bochníček and Vávra 2000, 2001) and also considering the mechanism of the ITZ formation (Bochníček and Vávra 2000) we can assume that at the barrier interfaces the Fe enrichment occurs as a consequence of outdiffusing of Si. As it was shown in our previous work (Vávra et al 2002), the ferromagnetic state of the $Fe_{0.5}Si_{0.5}$ thin film is retained up to a barrier thickness of about 0.7 nm (Mühge et al 1997). Based on these results, the Fe enrichment at the interfaces takes place only for a few atomic planes.

Taking this into account, the real structure of the junction probably should be considered as Nb/Fe/FeSi'/Fe/Nb. The thickness of the Fe layers is about one or two monolayers, and the FeSi' layer is the rest of the $Fe_{0.5}Si_{0.5}$ alloy. The particular stacking sequence mentioned above should be treated as the S/F/I/F/S junction. The existence of an insulating layer I (FeSi') in the structure is supported by a low value of the normal conductance $G_n \approx 0.2$ $\Omega^{-1}$. In comparison with our junction, the use of metallic barrier (e.g. CuNi (Veretennikov et al 2000), PdNi (Kontos et al 2001)) gives a few orders of magnitude higher normal conductance.

### 3.2. S/F/I/S junctions

To investigate in more details the effects mentioned above, a Nb/Fe$_{0.5}$Si$_{0.5}$/Si/Nb junctions were prepared by the same technology as Nb/Fe$_{0.5}$Si$_{0.5}$/Nb junctions. This type of junctions contains only one S/F interface which is like the bottom one in the Nb/Fe$_{0.5}$Si$_{0.5}$/Nb junctions.

Current-voltage curve of the Nb/Fe$_{0.5}$Si$_{0.5}$/Si/Nb junction is shown in fig. 6 and the experimental evidence of ac Josephson effect is clearly seen in the inset of fig. 7. In the differential conductance dI/dV vs. bias voltage curve (Fig. 7) the double enhancement of $G_n$ (arrow in fig. 7) is observed ($A \approx 2$). This enhancement is attributed to Andreev reflection at the bottom S/F interface. It is the same value of the Andreev reflection efficiency as it was observed in the case of the bottom interface, i.e. FeSi-on-Nb, of Nb/Fe$_{0.5}$Si$_{0.5}$/Nb junctions presented above. The peaks in the dI/dV curve with enhancement larger than two ($A > 2$) at the voltage $V < \Delta/e$ and the double enhancement ($A = 2$) at the voltage $V > \Delta/e$ are caused by the presence of the Si layer in the Nb/Fe$_{0.5}$Si$_{0.5}$/Si/Nb junctions.

Taking into account the Si outdiffusing from the Fe$_{0.5}$Si$_{0.5}$ layer, formation of only one S/F interface, and the observation of Andreev reflection at this S/F interface, the real structure of the Nb/Fe$_{0.5}$Si$_{0.5}$/Si/Nb junctions should be described as Nb/Fe/FeSi'/Si/Nb. The peak with enhancement $A > 2$ for the $V < 0.9\ mV$ is caused by the fact that low energy electrons coming from the Nb layer to the Fe layer can be reflected back at the Fe/FeSi' interface (Zareyan et al 2001). After that reflection the electrons can be Andreev-reflected again and due to that the differential conductance can be enhanced by the factor $A > 2$. This type of non-symmetric junctions consists of one Si barrier and one S/F interface connected in series. The voltage on the barrier is added to the voltage attributed to the S/F interface, thus the Andreev reflection occurs at the junction voltages $V > \Delta/e$, too.

### 4. Conclusion

In conclusion, in our S/F/I/F/S Josephson junctions with thin F layers, almost quadruple enhancement of the differential conductance, compared with that of the normal conductance, was observed as the consequence of Andreev reflection. To obtain a high value of transparency T, a low roughness of S/F interface and uniform concentration of ferromagnetic material along the S/F interface are needed (Zareyan et al 2001). We have shown that the transparency of the S/F interfaces in the Nb/Fe/FeSi'/Fe/Nb structures is close to unity, which is almost ideal. This is the consequence of a good layering quality of the S/F interfaces as well as of a uniform Fe concentration along the interfaces. These results are also supported by the observation of double enhancement of the differential conductance compared with that of the normal conductance in the S/F/I/S Josephson junctions. The observed characteristics as well as the efficiency of Andreev reflection $A$ in both types of junctions are well reproducible.

This work is supported by the VEGA grant Agency, projects No. 459/0214/2003 and 2/3116/23 and by the GOA/2004/02 and ESF VORTEX Programs.

**Figure Captions:**

**Fig. 1.** A cross-sectional high resolution TEM image of the Nb/Fe$_{0.5}$Si$_{0.5}$/Nb tunnel junction. The amorphous structure of Fe$_{0.5}$Si$_{0.5}$ barrier layer is clearly seen. No crystalline bridges between Nb layers are observed.

**Fig. 2.** Temperature dependence of the zero-bias resistance of the Nb/Fe$_{0.5}$Si$_{0.5}$/Nb junction R$_{TJ}$(T) confirms its tunnel-like character.

**Fig. 3.** Current-voltage characteristics of the Nb/Fe$_{0.5}$Si$_{0.5}$/Nb superconducting tunnel junctions at 4.2 K. Inset: Zoom of the current-voltage characteristics of such junction (a) without microwave irradiation (shifted by 20 µA for better clarity) and (b) with irradiation by 10 GHz. The Shapiro steps confirm the presence of the ac Josephson effect.

**Fig. 4.** Differential conductance vs. bias voltage of Nb/Fe$_{0.5}$Si$_{0.5}$/Nb junction at 4.2 K, showing a quadruple enhancement of the conductance compared with that of the normal state. Since in the superconducting state (dI/dV(V=0) is infinity), the dI/dV curve is shown for voltages V ≠ 0.

**Fig. 5.** Temperature evolution of the differential conductance characteristics for a temperature range **(a)** from 4.23 to 5.03 K and **(b)** from 5.54 to 6.12 K.

**Fig. 6.** Current-voltage characteristics of the Nb/Fe$_{0.5}$Si$_{0.5}$/Si/Nb superconducting tunnel junctions at 4.2 K. Inset: Zoom of current voltage characteristics of such junction (a) without microwave irradiation and (b) with irradiation by 10 GHz. The Shapiro steps confirm the presence of the ac Josephson effect.

**Fig. 7.** Differential conductance vs. bias voltage of the Nb/Fe$_{0.5}$Si$_{0.5}$/Si/Nb junction at 4.2 K, showing a double enhancement of the conductance compared with that of the normal state.

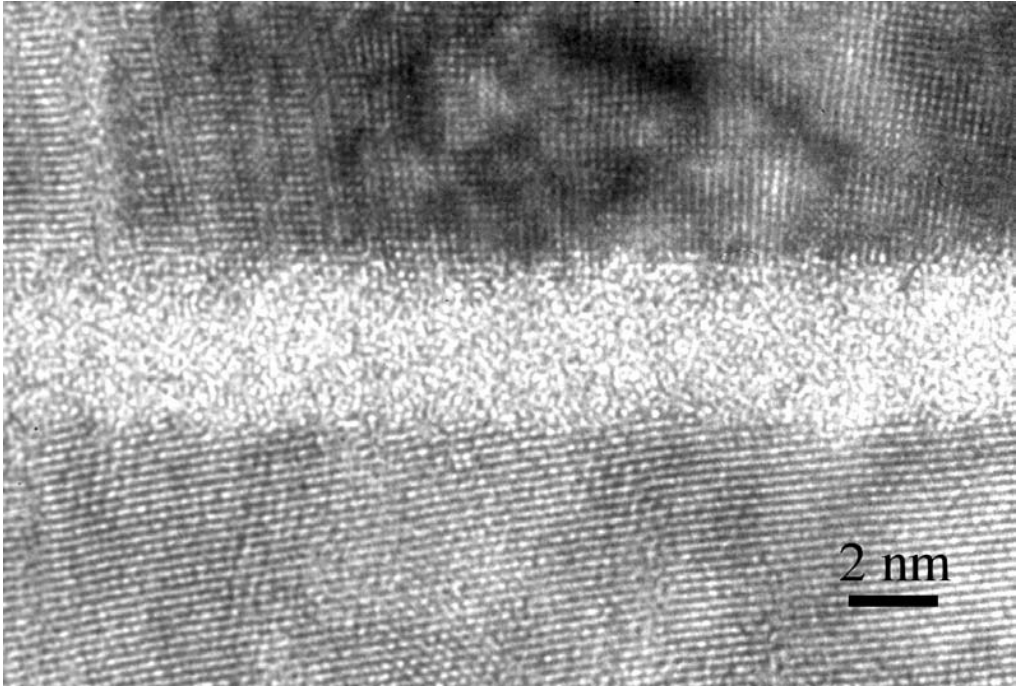

**Fig. 1.**

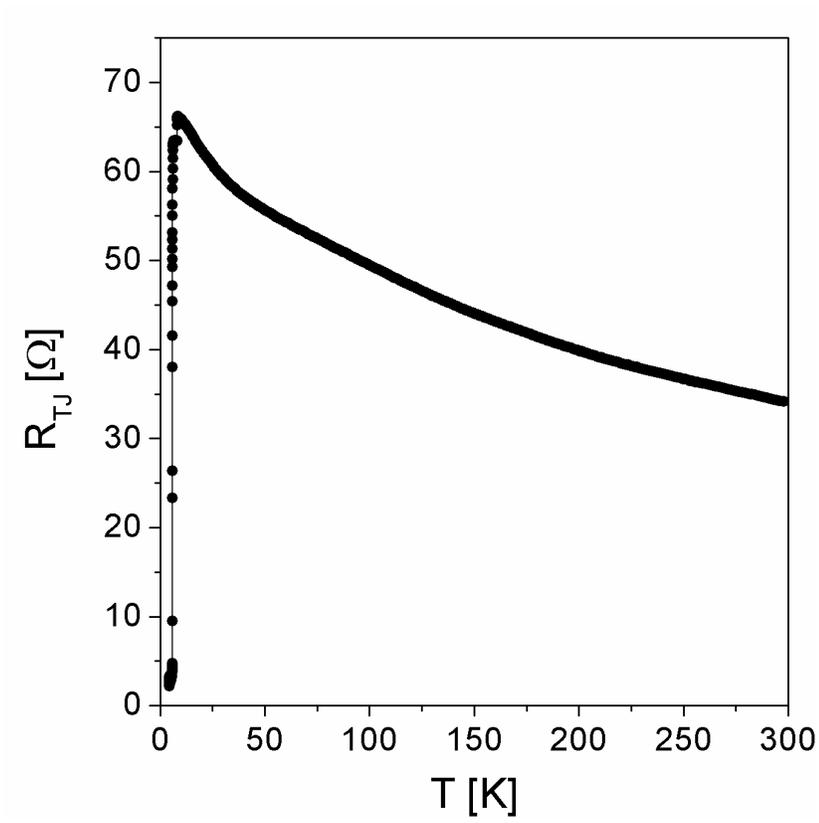

**Fig. 2.**

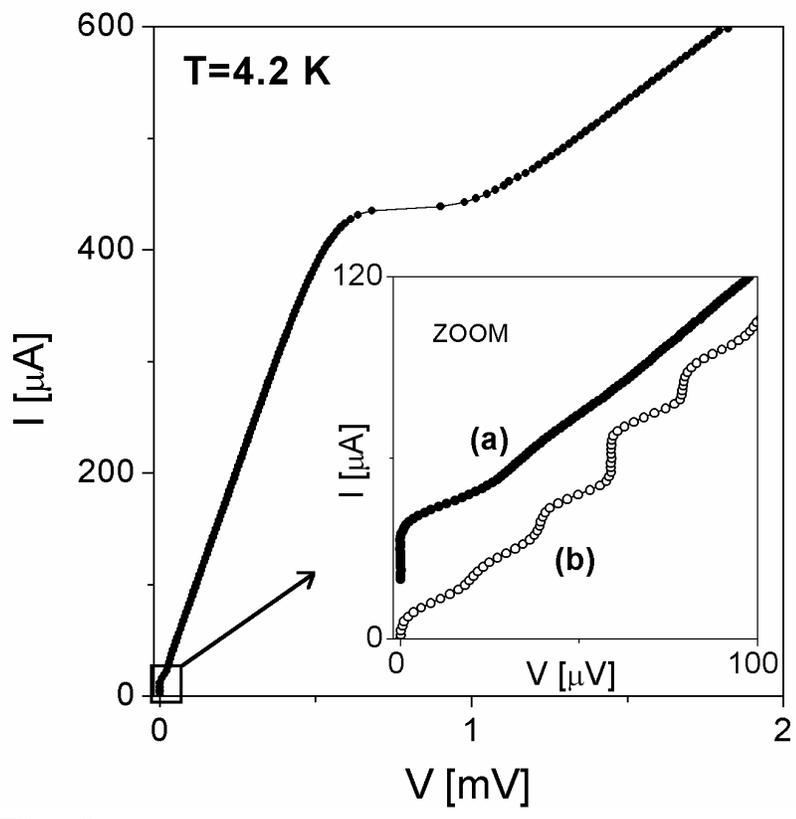

**Fig. 3.**

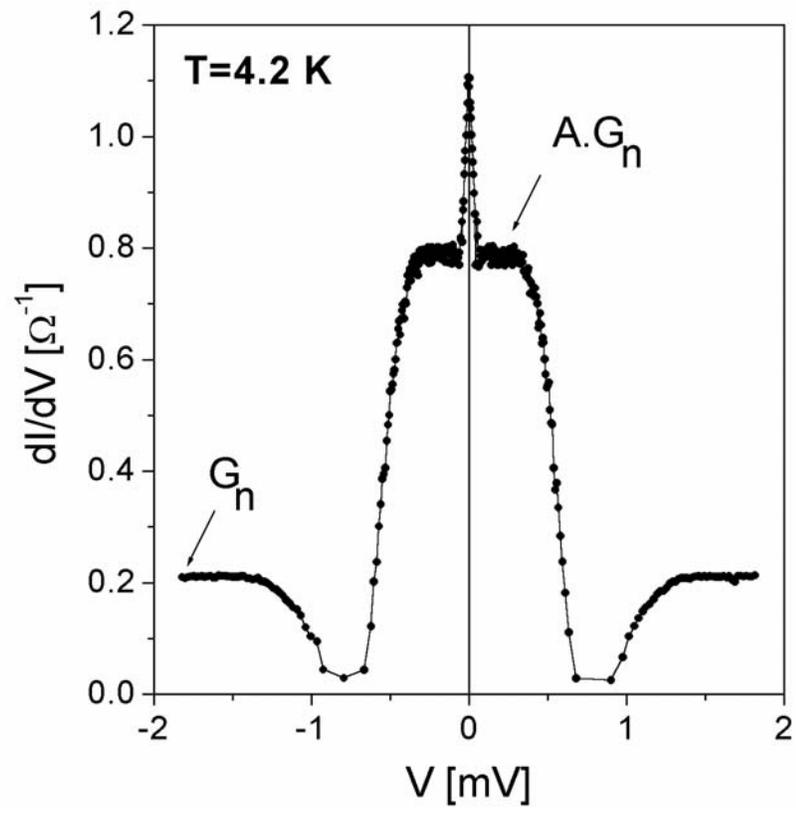

**Fig. 4.**

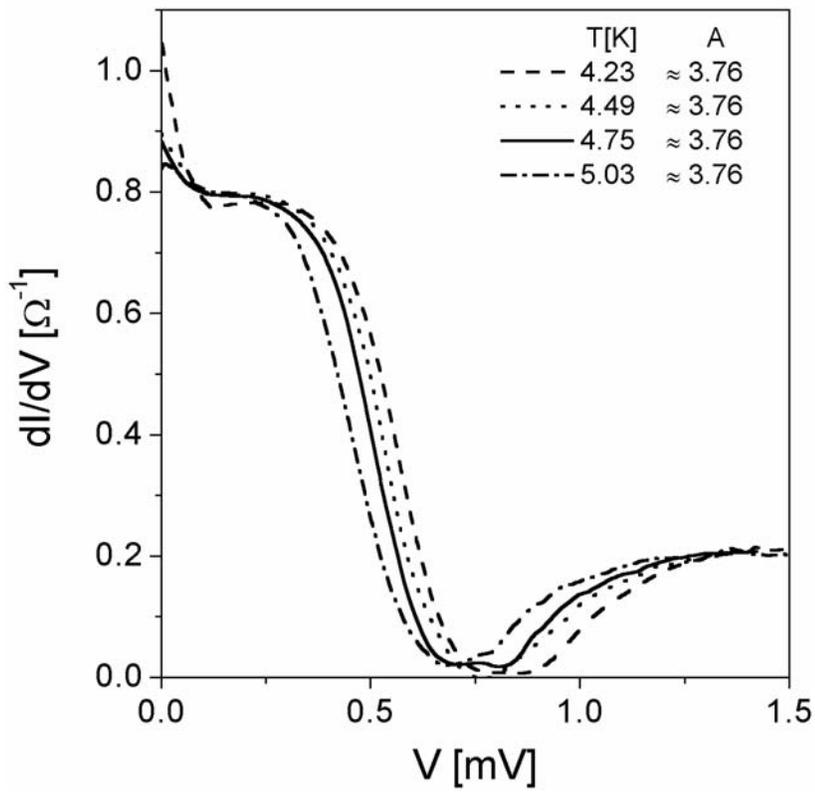

**Fig. 5a.**

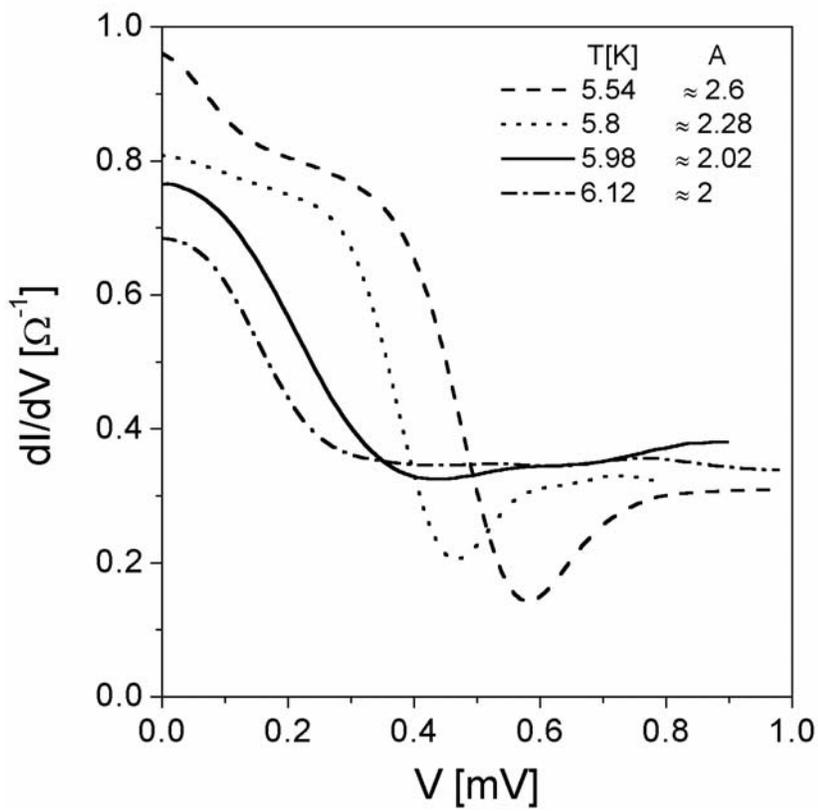

**Fig. 5b.**

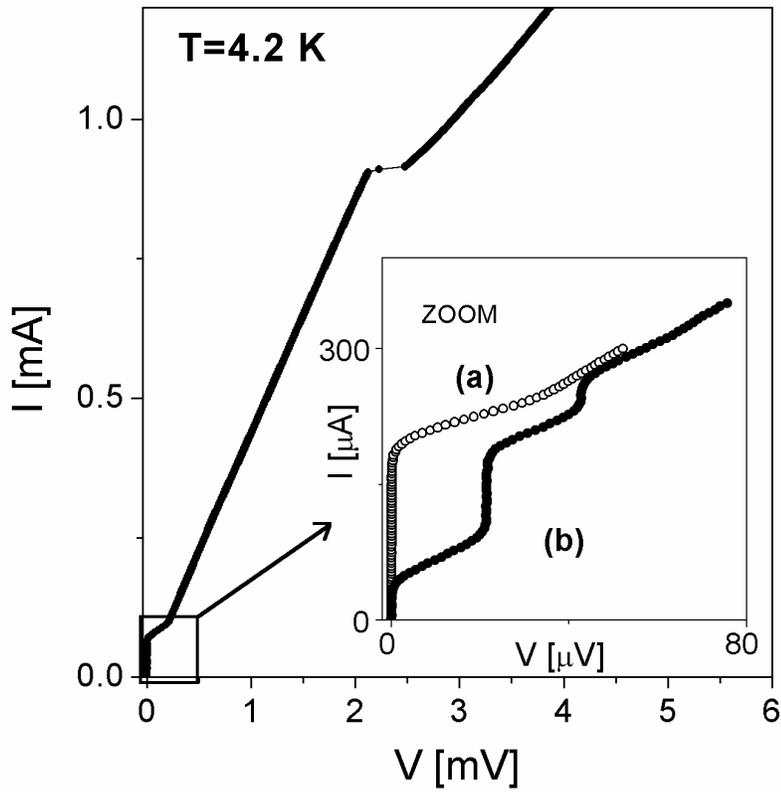

**Fig. 6.**

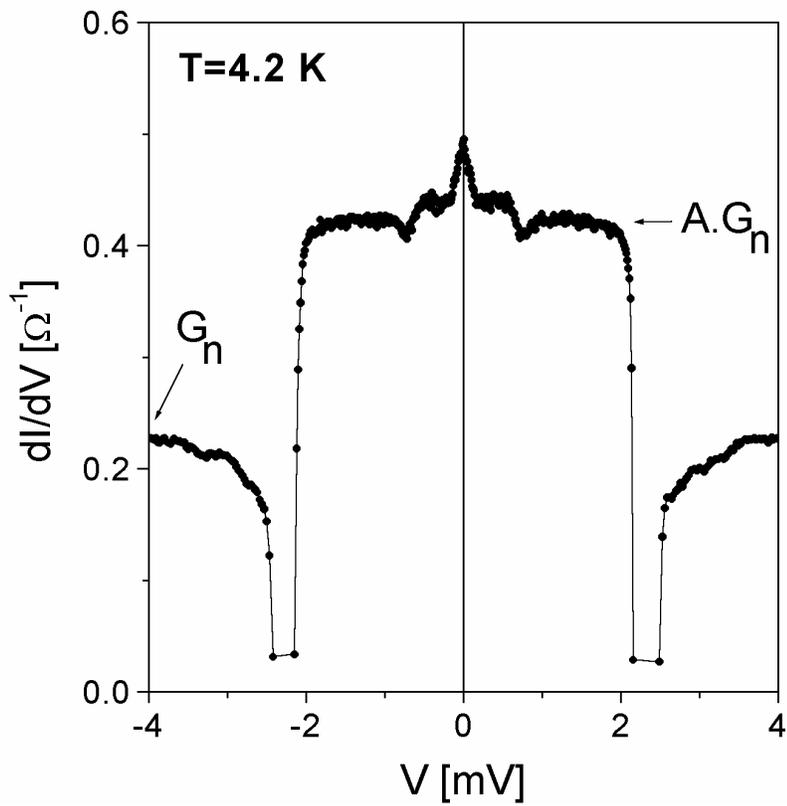

**Fig. 7.**

15